\documentclass[10pt,onecolumn]{emulateapj}

\newcommand{\be}{\begin{eqnarray}} 
\newcommand{\ee}{\end{eqnarray}} 
\newcommand{\Msol}{\mbox{$M_{\odot}\;$}}
\newcommand{\Msun}{\mbox{$M_{\odot}\;$}}
\newcommand{\tcmin}{$T_c^{\mathrm{min}}$}
\newcommand{\tcmax}{$T_c^{\mathrm{max}}$}

\newcommand\simgreater{\buildrel > \over \sim}
\newcommand\simless{\buildrel < \over \sim}

\begin{document}

\title{Neutrino Emission from Cooper Pairs and Minimal Cooling of Neutron Stars}

\author{Dany Page}
\affil{Departamento de Astrof\'{\i}sica Te\'orica, 
      Instituto de Astronom\'{\i}a,
      Universidad Nacional Aut\'onoma de M\'exico, 
      04510 Mexico D.F., Mexico}
\email{page@astro.unam.mx}
\author{James M. Lattimer}
\affil{Department of Physics and Astronomy, State University of New York 
at Stony Brook, Stony Brook, NY-11794-3800, USA}
\email{lattimer@astro.sunysb.edu}
\author{Madappa Prakash}
\affil{Department of Physics and Astrononmy, Ohio University, Athens,
 OH 45701-2979, USA}
\email{prakash@harsha.phy.ohiou.edu}
\author{Andrew W. Steiner}
\affil{Joint Institute for Nuclear Astrophysics, National
 Superconducting Cyclotron Laboratory and, \\ Department of Physics
 and Astrononmy, Michigan State University, East Lansing, MI 48824, USA}
\email{steinera@pa.msu.edu}
\shorttitle{COOLING OF NEUTRON STARS}
\shortauthors{PAGE, LATTIMER, PRAKASH, \& STEINER}

\begin{abstract}

The minimal cooling paradigm for neutron star cooling assumes that
enhanced cooling due to neutrino emission from any direct Urca
process, due either to nucleons or to exotica such as hyperons, Bose
condensates, or deconfined quarks, does not occur.  This scenario was
developed to replace and extend the so-called standard cooling
scenario to include neutrino emission from the Cooper pair breaking
and formation processes that occur near the critical temperature for
superfluid/superconductor pairing.  Superfluidity is generally
expected to exist in the neutron star interior, and Cooper-pair
neutrino emission processes, which operate through both vector and
axial channels, can dominate cooling in the minimal model.  Neutron
stars that have observed temperatures that are too low for their age
than in the minimal cooling model for any combination of its
parameters will imply that enhanced cooling is occurring.  Previous
studies showed that the observed temperatures of
young, cooling, isolated neutron stars with ages between $10^2$ and
$10^5$ years, with the possible exception of the pulsar in the
supernova remnant CTA 1, are consistent with predictions of the
minimal cooling paradigm as long as the neutron $^3$P$_2$
pairing gap present in the stellar core is of moderate size.

Recently, it has been found that Cooper-pair neutrino emission from
the vector channel is suppressed by a large factor, of order
$10^{-3}$, compared to the original estimates that violated vector
current conservation.  We show that Cooper-pair neutrino emission
remains, nevertheless, an efficient cooling mechanism through the
axial channel.  As a result, the elimination of neutrino emission from
Cooper-paired nucleons through the vector channel has only minor
effects on the long-term cooling of neutron stars within the minimal
cooling paradigm.  We further quantify precisely the effect of the
size of the neutron $^3$P$_2$ gap and demonstrate that 
consistency between observations and the minimal cooling paradigm
requires that the critical temperature $T_c$ for this gap covers a range of
values between $T_c^{\mathrm{min}} \simless 0.2 \times 10^9$ K
up to $T_c^{\mathrm{max}} \simgreater 0.5 \times 10^9$ K in the core of the star.
This range of values guarantees that the Cooper-pair neutrino emission
is operating efficiently in stars with ages between $10^3$ to $10^5$ years,
leading to the coldest predicted temperatures for young neutron stars.
In addition, it is required
that young neutron stars have heterogenous envelope compositions: some
must have light-element compositions and others must have
heavy-element compositions.  Unless these two conditions are
fulfilled, about half of the observed young cooling neutron stars are
inconsistent with the minimal cooling paradigm and provide evidence
for the existence of enhanced cooling.

\end{abstract}


\section{INTRODUCTION}

Neutron stars older than several minutes cool through a combination of
emission of neutrinos from their interiors and photons from their
surfaces.  The former dominates the energy losses during the first
$\sim 10^5$ years after birth, but the latter is responsible for the
observed thermal emissions detected from several neutron stars older
than hundreds of years.  For neutron stars of this age and older, the
effective surface temperatures are tightly coupled to the interior
temperatures, and the observability of these stars therefore depends
crucially on the overall rate of neutrino emissions from within the
star.  

Historically, theoretical neutron star cooling models have fallen into
two categories, ``standard" cooling or enhanced cooling (see,
e.g. \citealt{P92}, \citealt{Pr94,Pr98}, \citealt{Petal06}, and \citealt{P09}). 
Standard cooling implies
that no enhanced neutrino emissions from any direct Urca processes
(\citealt{LPPH91} and \citealt{PPLP92}), due either to nucleons or to
exotica such as hyperons, Bose condensates or deconfined quarks,
occurs.  \cite{PLPS04} (hereafter referred to as Paper I) introduced
the minimal cooling paradigm to address the question of whether or not
there exists observational evidence that firmly indicates that
enhanced cooling takes place in some neutron stars.

Traditionally, the standard cooling scenario has included only the
modified Urca process \citep{FM79,Tsur86}.  However, over the last
decade it has been realized that another important cooling mechanism
is provided by Cooper pair breaking and formation~\citep{FRS76,VS87},
termed as the ``PBF'' process, which occurs in the presence of
superconductivity or superfluidity in dense matter.  It is generally
accepted that superfluidity occurs in neutron star matter, although
the magnitudes of the superfluid gap energies as a function of density
are still uncertain at present.  Furthermore, for temperatures near
the associated superfluid critical temperatures, emission from Cooper
pairs dominates the neutrino emissivities in many cases.  Therefore,
in the minimal cooling paradigm, neutrino emission is included both
from modified Urca and bremsstrahlung processes as well as from the
Cooper pair breaking and formation processes.  Enhanced cooling is
assumed not to occur.  Any neutron star that has an observed surface
temperature too low for its age than that predicted by the minimal
cooling model for any combination of its input ingredients therefore
implies that enhanced cooling as defined above has occured in that
star.

Using the rates for Cooper-pair processes established by previous
authors \citep{YKL99, KHY99}, Paper I concluded that with the possible
exception of the pulsar in 3C58 {\footnote{In view of the recent
   revisions in its age, the luminosity of this object now lies close
   to the cooling curves of the minimal paradigm.}} and CTA 1, all
cooling neutron stars for which thermal emissions have been detected
are marginally consistent with the assumption that enhanced cooling is
absent, given the combined uncertainties in ages and temperatures or
luminosities. 
Importantly, it was concluded that overall consistency 
with minimal cooling was possible only if the neutron
$^3$P$_2$ gap was similar to our model "a" in Paper I,
i.e., with critical temperatures of the order of $10^9$ K.

This result was not sensitive
to the neutron star mass or to the sizes of the n and p $^1$S$_0$
gaps. In the case that the n $^3$P$_2$ gap was significantly larger
than this value, theoretical models yielded too slow cooling to
account for the low temperatures of about half of the young neutron
stars so far observed in thermal emission. It is also apparent from
the results of Paper I that overall consistency with minimal cooling
requires heterogeneity in envelope compositions for the young stars:
light element compositions for some and heavy element compositions for
others. One of the purposes of this paper is to show that, besides not
being too large, the n $^3$P$_2$ gap should also be not too small.
Unless the n $^3$P$_2$ lies in a narrow range, and there is
heterogeneity in young neutron star envelopes, the implication is that
about half of the observed young neutron stars with thermal emission
have some degree of enhanced cooling.

Of course, even if the conditions on the neutron $^3$P$_2$ gap and
envelope compositions are met, the overall consistency of the minimal
cooling paradigm with observations does not necessarily mean that
direct Urca processes are forbidden.  Direct Urca processes may in
fact still occur in many stars, but those neutron stars would quickly
grow too cold to be detectable from their thermal emissions
\citep{PA92}.

Pairing is expected to appear at low Fermi momenta ($k_F\simless 1.5$
fm$^{-1}$) in the singlet $^1$S$_0$ state \citep{BMP58} and in the triplet
$^3$P$_2$-$^3$F$_2$ mixed channel at higher momenta \citep{T72}. 
In both cases, singlet
or triplet, Cooper-pair emission can occur through the vector or the
axial channel. 
The conservation of vector current in dense superfluid neutron matter
was first addressed by \cite{KR04}.
Recently, \cite{LP06a} have demonstrated that the
vector part of the Cooper-pair emission in a one-component system of
paired fermions is suppressed by a factor of $\sim (1/20)(v_{F}/c)^4
\sim 10^{-3}$ relative to the original estimates of \cite{FRS76},
where $v_F$ is the velocity of particles at the Fermi surface and $c$
the speed of light. \cite{LP06a} have identifed the key reason for
this large suppression in one-component matter: in their original work,
\cite{FRS76} employed the bare vertex in the vector weak current;
however, the use of the bare vertex violates vector current
conservation. The conservation of vector current is achieved by
including collective effects which minimally requires (in order to
satisfy the Ward identity) that the correction to the bare vertex is
calculated to the same order of approximation as the quasi-particle
propagator is modified by the pairing interaction in the system.

The large suppression in the vector channel has been confirmed by
additional works~\citep{SMS07,EV08,L08,SR08}.  If the total
Cooper-pair emission were suppressed by this factor, significant
changes to results of the minimal cooling model might be
anticipated. However, Cooper-pair emission occurs through both vector
and axial channels, and the axial part of the Cooper-pair emission,
which is proportional to $v_{F}^3$, is only slightly modified. Also,
the rate of Cooper pair emission from the triplet configuration, being
due to axial currents, is largely unaffected by these new results.  A
cursory examination of the results of Paper I suggests additional
reasons that the revised Cooper-pair rates might not have a major
influence on the minimal cooling scenario.  One reason is that the
specific heat of matter with superfluidity is not affected by the
existence or absence of Cooper-pair emission.  As a result, the
cooling associated with $^1$S$_0$ neutrons, which is confined to the
crust, has only a transitory effect that is important during the
crust's thermal relaxation (i.e., during the first few hundred years
at most).  Another reason is that the cooling associated with the
$^1$S$_0$ protons in the stellar core proceeds predominantly through
the axial channel.  Nevertheless, the large magnitude of the vector
suppression suggests that a quantitative re-analysis of the cooling is
in order to confirm these expectations. 

The purposes of this paper are to (i) assess the effects of the
suppression of Cooper pair neutrino emission in the vector channel on
neutron star cooling and to examine their effects on the minimal
cooling paradigm, (ii) critically evaluate the extent to which current
data are compatible with the minimal cooling paradigm, (iii) highlight
the crucial role of triplet pairing (found in Paper I) in more
quantitative terms, and (iv) emphasize that young neutron stars cannot
have identical envelope compositions and remain compatible with the
minimal cooling paradigm.

In presenting our results below, the key ingredients of the minimal
model in Paper I are retained.  In doing so, we utilize the same (i)
equation of state, (ii) superfluid properties of the relevant
components, (iii) envelope composition, and (iv) stellar mass as in
Paper I.  We note that in a neutron star crust, which is a
multi-component system, the neutron PBF emissivity might be modified
by the lattice~\citep{SR08}.  However, this effect has not yet been
computed in detail. For the purposes of this work, we will assume that
the modification in the speed of sound is small, and, thus, the
$v_{F}^4$ suppression in the vector channel for homogeneous bulk
matter holds in general.  The neutrino emissivities from the PBF
processes will be altered to those calculated recently so that an
apposite comparison with Paper I can be made. Where appropriate, new
inputs with regard to superfluid gaps and revisons in data from
observations will be utilized and so indicated.

\section{PBF NEUTRINO EMISSIVITIES}

As the temperature nears the  critical temperature $T_c$ for pairing,
new channels for neutrino emission through the continuous formation
and breaking of Cooper pairs \citep{FRS76,VS87} become operative.

The PBF emissivity for both neutrons ($i=n$) and
protons ($i=p$) in the singlet ($j=s$) and triplet ($j=t$) channels
(we only consider $m_J=0$) can be written in the form~\citep{FRS76,YKL99}
\begin{equation}
Q = \frac{4 G_F^2 m_{i}^{*} p_{F,i}}{15 \pi^5 \hbar^{10} c^6}
\left(k_B T\right)^{7} {\cal N}_{\nu} a_{i,j} F_j\left[\Delta_i(T)/T\right] \,,
\end{equation}
where $G_F$ is the Fermi weak-interaction constant, $m_i^*$ are 
effective masses, $p_{F,i}$ are Fermi momenta, $T$ is the temperature,
${\cal N_\nu}$ is the number of neutrino flavors, $a_{i,j}$ are factors
involving vector and axial coupling constants and $F_j$ are functions
that depend on the temperature dependent pairing gaps $\Delta_i(T)$ and
control the efficiency of the PBF process.   The functional forms of
the gaps $\Delta_i(T)$ are discussed in \S 3.3 of Paper I.
Taking ${\cal N}_{\nu} = 3$
\begin{eqnarray}
Q &=& 3.51\times 10^{21}~
\frac{\mathrm{erg}}{\mathrm{cm}^3~\mathrm{s}}~
\left(\frac{m_i^{*}}{m_i}\right)
\left(\frac{p_{F,i}}{m_i c}\right) \times      
T_9^7 a_{i,j} F_j\left[\Delta_i(T)/T\right] \, ,
\label{q}
\end{eqnarray}
where $m_i$ are the bare masses , and 
the control functions $F_s$ and $F_t$ (for $m_J=0$) are
\citep{YKL99}
\begin{eqnarray}
F_s &=& y^2 \int_0^{\infty} \frac{z^4~dx}{\left(1+e^z\right)^2} \\
F_t &=& \frac{1}{4 \pi} \int 
d \Omega~y^2 \int_0^{\infty}
\frac{z^4~dx}{\left(1+e^z\right)^2} 
\end{eqnarray}
where $y=\Delta_i(T)/T$, $z=\sqrt{x^2+y^2}$, and $\int d \Omega$
represents the angle averaging procedure detailed in \citet{YKL99}.
The form of the control functions $F_s$ and $F_t$ shown
in Figure 13 of Paper I clearly shows that the PBF
process turns on when $T$ reaches $T_c$, increases its efficiency
as $T$ decreases, and becomes exponentially suppressed when the gap
approaches its maximum size $\Delta(T=0)$ when $T \simless 0.2 T_c$.

As we wish to examine the extent to which the cooling curves are altered 
by PBF emissivities that incorporate the conservation 
of the weak vector current, we collect below the relevant 
results for the factors $a_{i,j}$. 

{\em Vector current not conserved}: For the singlet and triplet
configurations, the factors $a_{i,j}$ (including both vector and axial
parts) have the values~\citep{YKL99,KHY99}
\begin{eqnarray}
a_{n,s} &=& C_{V,n}^2
+ C_{A,n}^2 ~\tilde{p}_{F,n}^2 \left(1+\frac{11}{42} \tilde{m}_n^{-2}
\right) \\
a_{p,s} &=& C_{V,p}^2 + C_{A,p}^2 ~\tilde{p}_{F,n}^2 
\left(1+
\frac{11}{42} \tilde{m}_p^{-2} \right) \\
a_{n,t} &=& C_{V,n}^2+2 C_{A,n}^2 \\ 
a_{p,t} &=& C_{V,p}^2+2 C_{A,p}^2 \,,
\end{eqnarray}
where $\tilde{p}_{F,i} \equiv p_{F,i}/m_i$ and 
$\tilde{m}_i \equiv m^*_i/m_i$.

{\em Vector current conserved}: Enforcing the Ward identity suppresses
the vector parts of the PBF emissivities~\citep{LP06a}. The
suppression in the vector part of the singlet channel is proportional
to $(4/81) v_{F,i}^4$~\citep{EV08}, where $v_{F,i}=p_{F,i}/m_i^* $ are
the Fermi velocities. The suppression of the vector part in the
triplet channel is also strong, so we set the contribution from the
vector part to zero as done in~\cite{LP06b}.  Consequently, the 
quantities $a_{i,j}$ are modified to
\begin{eqnarray}
a_{n,s} &=& C_{V,n}^2 \left(\frac{4}{81}\right) \left(\frac
{v_{F,n}}{c}\right)^4 + 
C_{A,n}^2 ~\tilde{p}_{F,n}^2 \left(1+\frac{11}{42} \tilde{m}_n^{-2}
\right) \\
a_{p,s} &=& C_{V,p}^2
\left(\frac{4}{81}\right) \left(\frac {v_{F,p}}{c}\right)^4 + 
C_{A,p}^2 ~\tilde{p}_{F,p}^2 
\left(1+
\frac{11}{42} \tilde{m}_p^{-2} \right) \\
a_{n,t} &=& 2 C_{A,n}^2 \qquad {\rm and} \qquad
a_{p,t} = 2 C_{A,p}^2 \,.
\end{eqnarray}
The values of the various coupling constants are $C_{V,n}=1$,
$C_{A,n}=g_A$, $C_{V,p}=4 \sin^2 \theta_W-1$, and $C_{A,p}=-g_A$,
where $g_A\approx 1.26$ and $\sin^2 \theta_W \approx 0.23$.
Notice that
$C_{V,p}^2 \ll C_{V,n}^2$, whereas $C_{A,p}^2 = C_{A,n}^2$.  The fact
that $\tilde{p}_{F,p} \ll \tilde{p}_{F,n}$, implies that neutrino
emission from the PBF process involving triplet neutron pairing is
intrinsically much more efficient than that from singlet proton
pairing.

Although the neutron pairing at high densities involves a
mixing of the $^3P_2$ and $^3F_2$ channels \citep{T72}, this mixing is
not taken into account in the triplet PBF emissivities quoted
above. In what follows, we will simply refer to the pairing of
neutrons at high densities as $^3P_2$ pairing.

\section{EFFECTS OF PAIRING AND PBF EMISSIVITIES
           \label{Sec:Coop}}

The most significant revision of PBF neutrino emission is in the case
of the neutron $^1$S$_0$ pairing as emission from the $^3$P$_2$ 
pairing is suppressed only by about 30\%.   Comparison of equations 
(5) through (8) with equations (9) through (12) shows that the total 
emissivity from the proton $^1$S$_0$ pairing is essentially unaffected 
as it is largely dominated by the axial channel.   Both proton 
$^1$S$_0$ and neutron $^3$P$_2$ pairings occur in the core, which 
contains more than 90\% of the star's volume, whereas the neutron 
$^1$S$_0$ pairing is essentially restricted to the crust.   We now
assess how the predicted large suppression in the vector
channel for the various gaps affects the minimal cooling paradigm.

\subsection{Effect of the neutron $^1$S$_0$ gap in the crust \label{Sec:crust}}

\begin{figure}
\begin{center}
\includegraphics[scale=0.30]{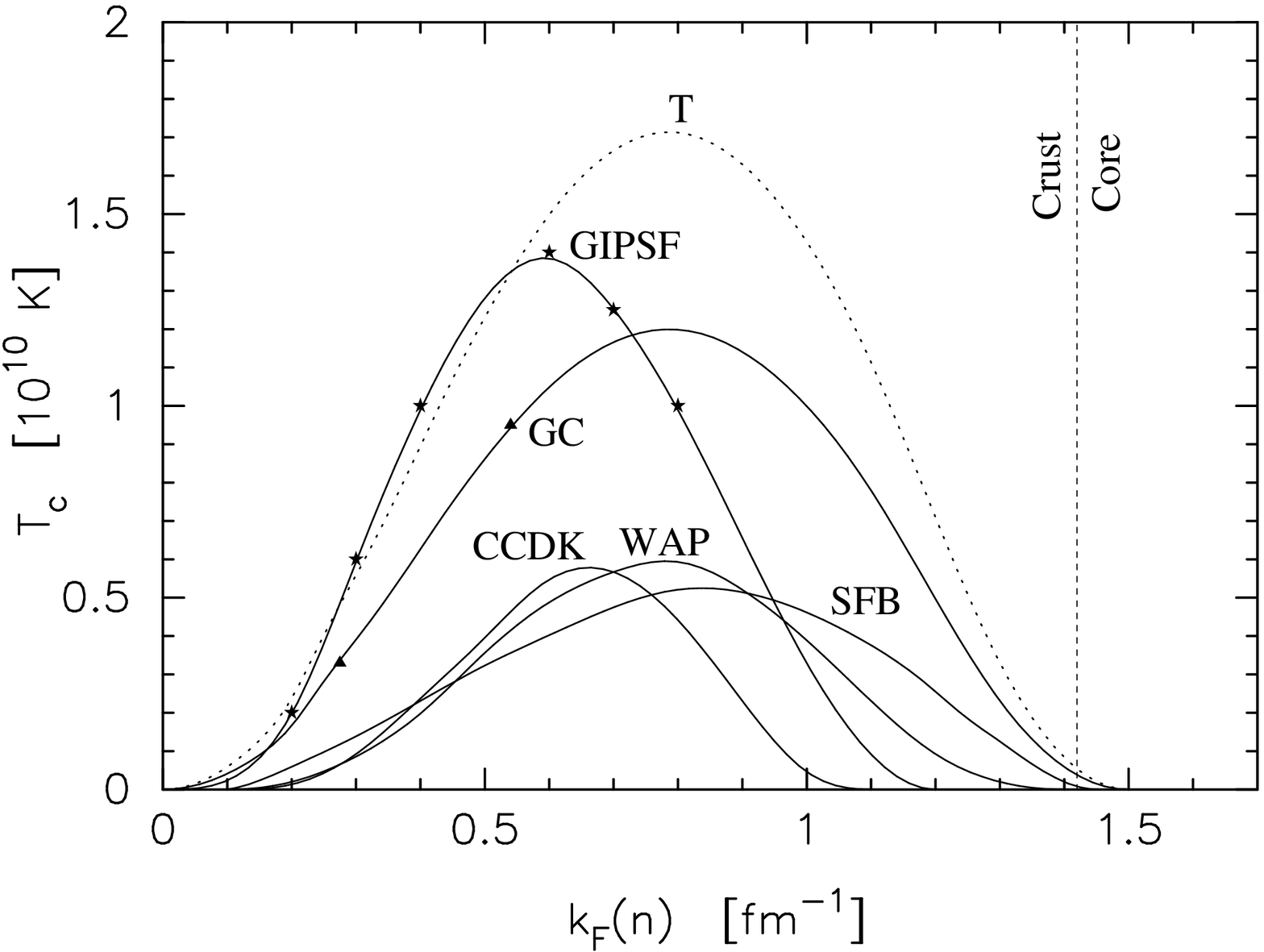}
\end{center}
\caption{Representative neutron $^1$S$_0$ pairing critical temperature
$T_c$ as a function of the neutron Fermi momentum $k_F$ from various
calculations: \cite{T72} (``T''), a calculation within the BCS
approximation; \citet{WAP93} (``WAP''), \citet{CCDK93} (``CCDK''), and
\cite{SFB03} (``SFB"), which take into account medium polarization
beyond the BCS level; \cite{GC08} (``GC") and \cite{Getal08}
(``GIPSF") denote Quantum Monte-Carlo results (as these latter
results, marked as triangles and stars, respectively, are restricted
to relatively low densities, we have extrapolated them to higher
densities).  In all cases, $T_c$ was obtained from the zero
temperature gap $\Delta(0)$ through the standard BCS result as $k_B T_c = 0.57
\Delta(0)$.  The vertical dotted line shows the location of the
crust-core boundary.
\label{Fig:n1S0}}
\end{figure}

\begin{figure}
\begin{center}
\includegraphics[scale=0.30]{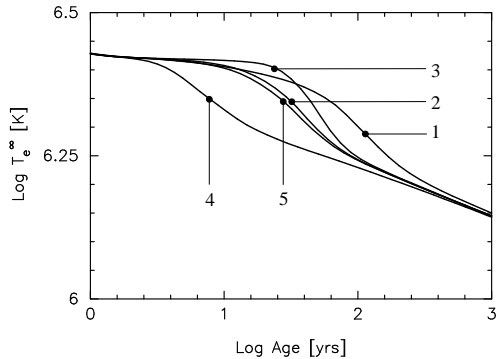}
\end{center}
\caption{Comparison of cooling during the crust thermal relaxation era
 including various effects from neutron $^1$S$_0$ pairing.  The
 $^1$S$_0$ gap ``SFB" from \cite{SFB03} has been employed. 
 No pairing in the core has been included and the star is a 1.4 \Msol star 
 built with the EOS of APR \citep{APR98} and has a heavy element envelope 
 (see Paper I).  Curves 1 through 5 are explained in the text.
\label{Fig1}}
\end{figure}

There are substantial variations in model predictions for the
$^1$S$_0$ gap. For the models we use, Figure~\ref{Fig:n1S0} summarizes
predictions for the superfluid critical temperatures $T_c$ as a
function of the neutron Fermi momentum $k_F(n)$, including results of
two new calculations performed after Paper I was written. The neutron
$^1$S$_0$ gap is mostly confined to the stellar crust, which
constitutes only a small fraction of the stellar volume. As a result,
effects of the suppression are mostly observed during the thermal
relaxation of the crust \citep{LvRPP94,P09}. We therefore focus on the
first $10^3$ years of evolution during which the effects of
suppression are most evident (see Figure~\ref{Fig1}).

\begin{figure}
\begin{center}
\includegraphics[scale=0.30]{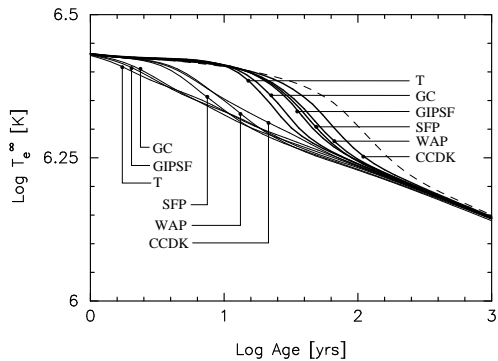}
\end{center}
 \caption{Early cooling of a neutron star with various neutron
 $^1$S$_0$ gaps. Upper (thick) curves take into account the
 suppression of the vector channel of the PBF process (as in curve 5
 of Figure 2) whereas, for comparison, the lower (thin) curves are
 without the supression (as in curve 4 of Figure 2).  
 The dashed curve show the cooling in absence of neutron pairing
 (as in curve 1 in Figure 2).
 Notation for gaps is the same as in Figure~\ref{Fig:n1S0}. }
\label{Fig2}
\end{figure}

We begin with a stellar model in which no pairing is taken into
account (curve 1 in Figure~\ref{Fig1}).  In this case, the crust's
early cooling is driven by neutrino emission from the plasmon 
process and the neutron-neutron bremsstrahlung from the 
(unpaired) neutrons in the inner crust, with a small contribution from 
the electron-ion (and a smaller one from electron-electron) 
bremsstrahlung process.

In the presence of a neutron $^1$S$_0$ gap, three effects appear:
suppression of the (inner crust) neutron specific heat, suppression of
the $n-n$ bremsstrahlung, and the onset of the PBF process.  We
illustrate the influence of these three effects in succession in
Figure~\ref{Fig1}.
Curve 2 includes only the suppression of the specific heat, which
results in a direct shortening of the thermal relaxation time of the
crust.  Curve 3 adds the $n-n$ bremsstrahlung suppression, which results
in a higher temperature and hence a lengthening of the relaxation phase.
Curves 4 and 5 show the total effect of the neutron $^1$S$_0$ pairing by
also taking into account the PBF emission process.  For comparison,
curve 4 is the result in which the vector channel contribution is
assumed to be unsuppressed whereas curve 5 takes this suppression into
account.  As had been anticipated, the suppression of the vector
channel of the PBF process has a significant effect, but only at
early times ($t<1000$ years), and results in warmer crusts and
increased crust relaxation times.  However, after crust relaxation is
attained, the differences among cooling histories of the various
models is very small (this is more clearly illustrated below in 
Figure~\ref{Fig:plot_data}).

In order to demonstrate that this result is independent of the
specific character of the $^1$S$_0$ gap, we compare in Figure~\ref{Fig2}
the cooling during the crust thermal relaxation era using
the various models for this gap from Figure~\ref{Fig:n1S0}.  Although
quantitative differences are apparent, the qualitative nature of the
effects of suppression are the same for all gaps.

The effects of vector suppression are therefore likely to be
important in the interpretation of neutron star crustal cooling
observed in X-ray transients from accretion-heated neutron stars in
Low-Mass X-ray Binaries \citep{Cacketal06,Cacketal08,Degetal09} as the
crustal cooling timescale is sensitive to the neutrino emission rates
\citep{LvRPP94,Rutetal02,Setal07,BC09}.  Inasmuch as the inclusion of
the effects of vector suppression increases the cooling timescale, the
inferred crust thickness will be overestimated if these effects are
ignored.  Likewise, higher temperatures can be reached in the crust of
an accreting neutron star once the vector suppression of the PBF rate
is taken into account and has important consequence for the triggering
of superbursts \citep{Cumetal06}.

We therefore conclude that the suppression of the vector channel of
the neutron $^1$S$_0$ PBF process does not lead to a distinguishable
effect in the long-term cooling ($>$ 1000 years) of the star.

\subsection{Effects of the neutron $^3$P$_2$ and proton $^1$S$_0$ gaps  
in the core}

\begin{figure}
\begin{center}
\includegraphics[scale=0.40]{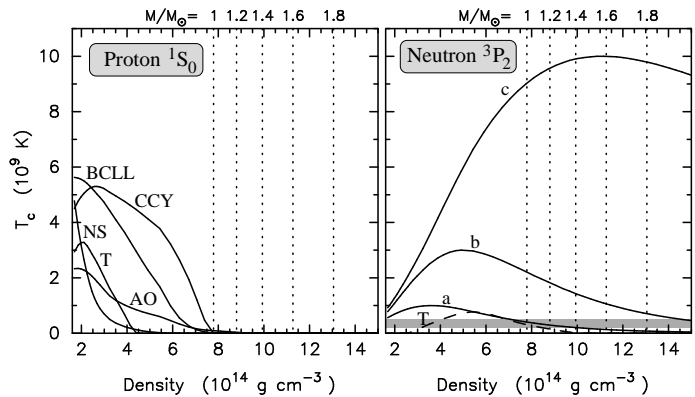}
\end{center} 
\caption{Critical temperature, $T_c$, for the proton $^1$S$_0$ gaps,
 left panel, and neutron $^3$P$_2$ gaps, used in this work. These are
 the same gaps as shown in Figure 9 and 10 of Paper I, respectively,
 but displayed as functions of the matter density $\rho$ instead of
 particle Fermi momenta, $k_F$, as in Paper I. Conversion from $k_F$
 to $\rho$ was performed with the APR EOS and its corresponding
 proton fraction. This conversion is only weakly dependent on the EOS
 given the constraints imposed by the conditions of minimal cooling
 on the EOS. Central densities of stars with masses from $1$ to $1.8$
 \Msun are indicated. The grey strip for the neutron $^3$P$_2$ gaps
 is the {\em compatibility band} discussed in \S \ref{Sec:3P2opt}.
 See Paper I for references.
\label{Fig:Tc}} 
\end{figure}

We now compare the relative efficiencies of the PBF processes from the
neutron $^3$P$_2$ and/or proton $^1$S$_0$ Cooper pairs with the
modified Urca processes (neutron and proton branches) and nucleon
bremsstrahlung processes in the core. In Figure \ref{Fig:Tc}, we plot
the two families of gaps we will consider. As discussed in Paper I,
these gaps cover the broad range of predicted values from microscopic
calculations.  Particularly uncertain is the maximum size of the
neutron $^3$P$_2$ gap, as well as the range in density over which it
is significant \citep{SF04,BEEHS98}.

\begin{figure}
\begin{center}
\includegraphics[scale=0.35]{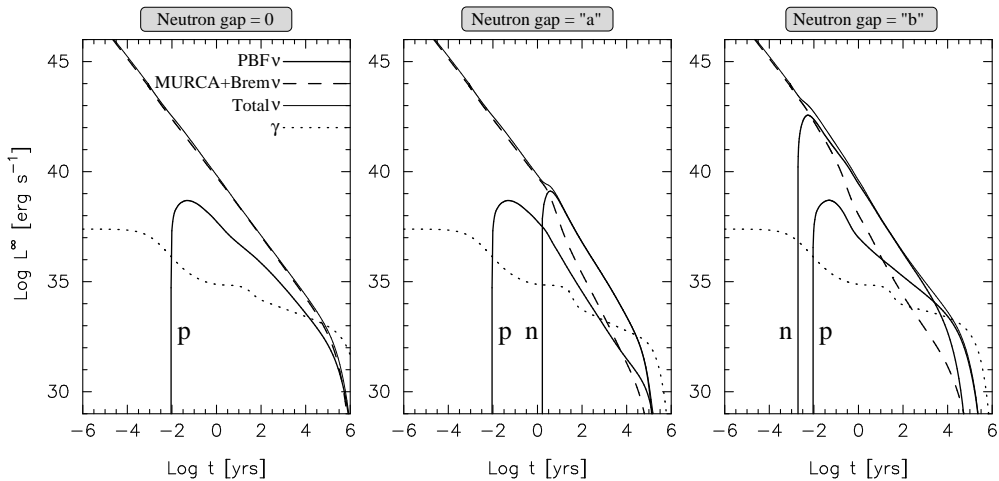}
\end{center} 
\caption{Comparison of luminosities from various processes during
three realistic cooling histories: photon (``$\gamma$''), all
$\nu$-processes (``Total $\nu$''), modified Urca and nucleon
bremsstrahlung (``MURCA+Brem $\nu$''), and PBF (``PBF $\nu$'') from
neutron $^3$P$_2$ (``n'') and proton $^1$S$_0$ (``p'') pairing.  PBF
neutrino emission from the neutron $^1$S$_0$ gap is not shown explicitly
as its contribution is always dominated by other processes, but is
included in the total $\nu$ luminosity.  Suppression of the vector
channel of the PBF processes is properly taken into account in all
cases.  In all cases shown, the proton $^1$S$_0$ gap is from
\cite{AO85a} (model ``AO" in Figure \ref{Fig:Tc}) and the neutron
$^1$S$_0$ gap from \cite{SFB03} (model ``SFB" in Figure~\ref{Fig:n1S0}).
The neutron $^3$P$_2$ gap is chosen to be vanishingly small (left
panel), from our model ``a'' (center), or model ``b'' (right) 
from Figure \ref{Fig:Tc}.  The star is a 1.4 \Msun star
built with the EOS of APR \citep{APR98} and has a heavy element
envelope (see Paper I).
\label{Fig:Cool_Lum}} 
\end{figure}

As an example, we choose the proton $^1$S$_0$ gap from model "AO" of
Figure \ref{Fig:Tc} and we consider three neutron $^3$P$_2$ gaps: one
vanishingly small, and the two models ``a'' and ``b'' from the same
figure, which have maximum values of $T_c=10^9$ K and $3\times10^9$ K,
respectively.  Figure~\ref{Fig:Cool_Lum} shows the resulting neutrino
and photon luminosities in full cooling calculations.  In the case of
a vanishing neutron $^3$P$_2$ gap (left panel), the only PBF process
occurring in the core is from the proton $^1$S$_0$ pairing, but due to
its intrinsically low efficiency it cannot compete with the modified
Urca process which is unsuppressed in the inner core where the proton
gap vanishes.  This model is very similar to the old ``standard
cooling" case.  The other two panels in Figure \ref{Fig:Cool_Lum} with
non-vanishing neutron $^3$P$_2$ gaps clearly show that the PBF
processes dominate the total neutrino luminosity as soon as the
neutron $^3$P$_2$ pairing appears.

It is worthwhile to note the competition between the proton $^1$S$_0$
and neutron $^3$P$_2$ PBF processes which depends on the relative
sizes of the gaps.  For a large neutron $^3$P$_2$ gap, as our case
``b" used to obtain the luminosities in the right panel of
Figure~\ref{Fig:Cool_Lum}, the temperature of the entire core drops
below $T_c$ in a short time; thereafter, the corresponding PBF process
is suppressed.  When neutrons in the entire core are well into the
superfluid phase the PBF process from the proton $^1$S$_0$ gap
subsequently drives the cooling, at ages $\simgreater 10^3$ yrs, but
with a low efficiency.  In contrast, when the neutron
$^3$P$_2$ model gap ``a" is used to obtain the luminosities in the
central panel of Figure~\ref{Fig:Cool_Lum}, neutrino emission from the
neutron PBF process largely dominates the cooling. As noted in Paper
I, such gaps as ``a" lead to the coldest minimal cooling neutron
stars.

\subsection{Characterization of the most efficient neutron 
$^3$P$_2$ gaps in the core\label{Sec:3P2opt}}

The most efficient pairing configurations, which lead to the coldest
neutron stars, a situation explored in Paper I, are neutron
$^3$P$_2$ gaps with $T_c$ values around $10^9$ K in the largest
possible fraction of the core (as in the case of our model ``a"). 
In this case, the efficient PBF process from the neutron $^3$P$_2$ gap
dominates the neutrino luminosity at ages $\sim 10^0 - 10^5$ yrs (as
seen in the central panel of Figure~\ref{Fig:Cool_Lum}) and results in
the coldest young neutron stars within the minimal cooling paradigm.

\begin{figure}
\begin{center}
\includegraphics[scale=0.30]{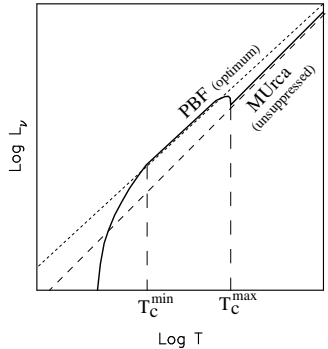}
\end{center} 
\caption{A schematic diagram of the neutrino luminosity as a function
 of temperature for the modified Urca and PBF processes. The dotted
 curve shows the optimal PBF luminosity (i.e., obtainable when a
 thick enough layer in the core has it temperature close to $T_c$)
 and the dashed curve shows the unsuppressed MUrca luminosity. The
 values of \tcmin~and \tcmax~are the minimum and maximum values of
 $T_c$ for the neutron $^3P_2$ gap (which is a function of density)
 that occur within the star. When the temperature in the core falls
 below \tcmax~the neutrino luminosity at that point increases to the
 PBF luminosity, which can be almost two orders of magnitude higher
 than the MUrca luminosity in the optimal case. When the temperature
 falls further, to below \tcmin, the neutrino luminosities from both
 the PBF and MUrca process are quenched. (Figure inspired from Figure
 20 of Paper I.)
\label{Fig:Cool_Coop}} 
\end{figure}

The schematic illustration in Figure~\ref{Fig:Cool_Coop} shows the
neutrino luminosity as a function of temperature for the modified Urca
and PBF processes. As long as the temperature is greater than
\tcmax, which we define as the maximum value of $T_c$ in the
core, the modified Urca process drives the cooling. When the
temperature falls below \tcmax, the PBF process turns on and
dominates the cooling, until the temperature drops below \tcmin,
which we define as the minimum value of $T_c$ in the core, when both
the PBF and modified Urca processes are quenched everywhere in the core.

\begin{figure}
\begin{center}
\includegraphics[scale=0.30]{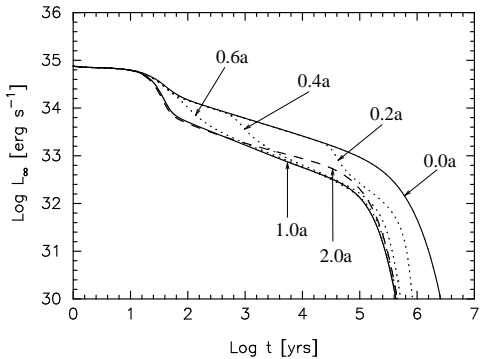}
\end{center} 
\caption{Comparison of predictions of the minimal cooling scenario for
 variations of the neutron $^3$P$_2$ gap.  We take as a reference gap
 our model ``a" for Figure \ref{Fig:Tc} which we scale by a factor
 ``s";  curves are labelled by this scale factor (the curve 0.0a is the
 gapless case and 1.0a represents the standard case ``a'').  The PBF
 process ensues when the cooling curve for each gap separates from the
 upper (gapless) trajectory.  The standard case 1.0a is seen to
 represent the most efficient neutrino emission and the most rapidly
 cooling case.  All models are for 1.4 \Msun stars built using the EOS
 of APR \citep{APR98} with heavy element compositions.  
\label{Fig:Cool_3P2}} 
\end{figure}

The surface temperature at early times is controlled by crustal
physics, as described in \S \ref{Sec:crust}, and is independent of the
evolution of the core. For the surface temperature to reach the
smallest possible values, the value of \tcmax~should be large
enough for the PBF process to turn on before, or not much later than, the
crust isothermalization time. A useful reference age is $\sim 10^3$
years, the estimated age of the youngest observed cooling neutron
stars, for the PBF process to be fully operating. At later times, if
the value of \tcmin~is too large both the PBF and modified Urca
processes will turn off before the photon cooling era and the cooling
will proceed at a slower pace. This feature is illustrated in
Figure~\ref{Fig:Cool_3P2} which shows cooling curves for various neutron
$^3$P$_2$ gaps. The upper solid curve shows cooling for the case 
in which the neutron $^3$P$_2$ gap is zero ($\Delta=0$). The lower
solid curve shows cooling for a neutron $^3$P$_2$ gap corresponding to
case ``a'', which is close to the most efficient case, giving the
lowest temperatures at all ages. The corresponding values of
\tcmax~and \tcmin~in the 1.4 \Msun star used in
Figure~\ref{Fig:Cool_3P2} are $10^9$ K and $2 \times 10^8$ K,
respectively. The three models ``0.6a", ``0.4a", and ``0.2a", in
Figure~\ref{Fig:Cool_3P2} with scaled down gaps show that to
obtain the coldest star \tcmax~should be at least about 0.5
times the \tcmax~value of model ``a", that is,
\be
T_c^{\mathrm{max}} > 0.5 \times 10^9~\mathrm{K}.
\label{Eq:Tcmax}
\ee
With respect to the optimal value of \tcmin, the model ``2.0a" in
Figure~\ref{Fig:Cool_3P2} with a scaled up gap shows that the turning
off of both modified Urca and PBF processes occurs somewhat too early,
resulting in warmer stars at ages $> 10^3$ years. Consequently,
\tcmin~should not be much larger than that of our model ``a"
for this particular 1.4 \Msun star:
\be
T_c^{\mathrm{min}} \simless 0.2 \times 10^9~\mathrm{K}.
\label{Eq:Tcmin}
\ee
The two bounds in equations \ref{Eq:Tcmax} and \ref{Eq:Tcmin} allow us
to draw a ``compatibility band" for the $T_c$ curve, shown as a
horizontal grey strip in the right panel of Figure \ref{Fig:Tc}. {\em
A $T_c$ curve will yield the coldest possible minimally cooling
neutron star, at ages between $10^3$ to a few times $10^4$ years, only
if it crosses the compatibility band within the density range present
in the core of the star.}  
Examination of Figure~\ref{Fig:Tc}
shows that neutron $^3$P$_2$ gaps resembling those of model ``a''
provide the most efficient cooling for neutron stars of masses close
to 1.4 solar masses.

In order to assess the role of the gap's density dependence, we also
performed cooling simulations using the neutron $^3$P$_2$ gap ``T''
(from \citealt{T72}) of Figure \ref{Fig:Tc}.  This gap model is
similar to our model ``a'', but the density dependence is different.
It has \tcmax = $0.75 \times 10^9$ K and \tcmin =0, this minimum being
reached at low densities, and thus crosses the compatibility band.
Cooling simulations utilizing models ``a'' and ``T'' result in
virtually identical trajectories.  In contradistinction to the lower
bound of equation~(\ref{Eq:Tcmax}), it is not possible to have an upper
bound on \tcmax, the only condition being that the $T_c$ curve must
cross the compatibility band.  
Notice that a gap with a density dependence similar to our models "b" and "c"
and a very large \tcmax ($\gg 0.5 \times 10^9$ K) is also likely to have a very large \tcmin,
and would violate the compatibility conditions in equations (\ref{Eq:Tcmax}) and (\ref{Eq:Tcmin})
for minimal cooling.
This density dependence, in particular an already large $T_c$ at nuclear matter density,
was inspired by the microscopic models of \cite{BEEHS98},
and the compatibility conditions we find are a strong constraint against such types of gaps.
However, in case the gap has a vanishingly small $T_c$ a low densities, as the model "T", 
and grows above $0.5 \times 10^9$ K at higher densities, then it will 
inescapably cross the compatibility band, no matter how large is its \tcmax.
It is intriguing that such strong constraints on the neutron
triplet gap emerge for consistency of the minimal model with data.

\section{DISCUSSION AND COMPARISON WITH DATA}

\begin{figure} 
\begin{center}
\includegraphics[scale=0.70]{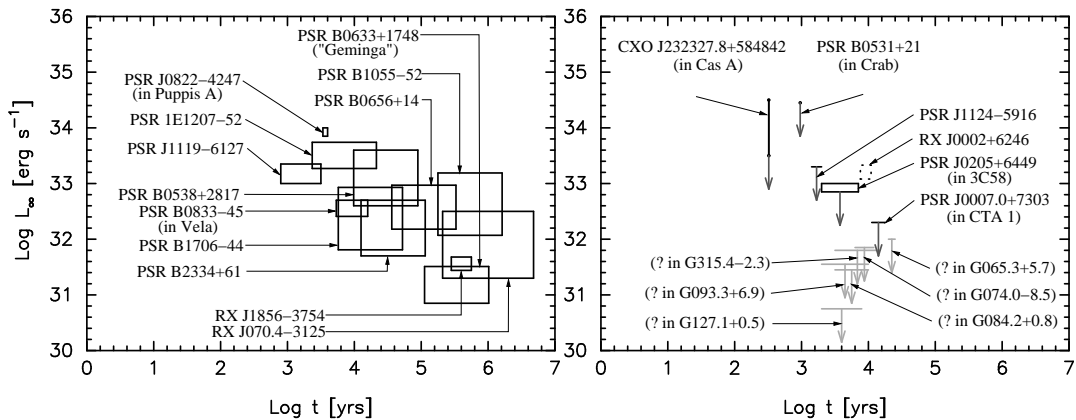} 
\end{center} 
\caption{Summary of observational data on thermal luminosities of
isolated cooling neutron stars.  
Left panel: 12 stars for which a thermal spectrum has been clearly detected.  
Right panel: the PSR in the nebula 3C58 seems to exhibit a thermal component,
and could be presented in the upper panel, whereas in the case of the other objects a
thermal component from the main stellar surface is not detected; consequently,
the data shown are upper limits on the thermal luminosity. 
The labels "?" indicate that a compact object has not yet been detected in
the supernova remnant.  Finally, the object RX J0002+6246 in CTB1 is
possibly not a neutron star \citep{Eetal08}.  See Appendix A for
details.  
\label{Fig:plot_data} } 
\end{figure}

\begin{figure} 
\begin{center}
\includegraphics[scale=0.70]{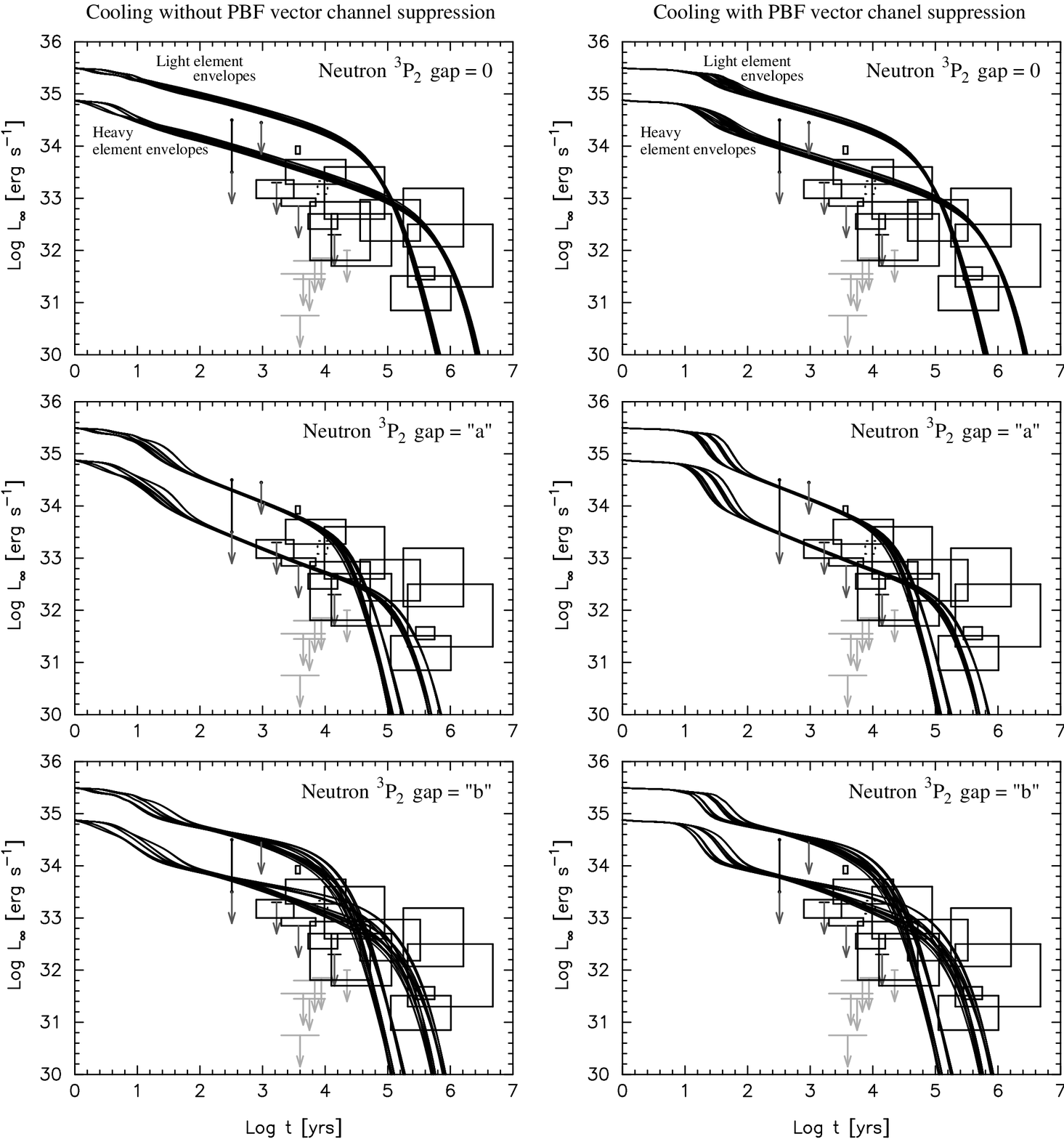} 
\end{center} 
\caption{Comparison of predictions of the minimal cooling scenario
with data; all models are for 1.4 \Msun stars built using the EOS of
APR \citep{APR98}. 
In the right panels the suppression of the
vector channel in the Cooper-pair neutrino emission is fully taken
into account whereas, for comparison, in the left panels the
supression has been omitted. 
In each row, the two panels have the same neutron $^3$P$_2$ gap, 
from a vanishing gap in the upper row
to our model gaps "a" and "b" (following the notations of Figure 10 in
Paper I) in the next two rows.  
In each panel two sets of cooling trajectories, either with light or 
with heavy element envelopes, are shown which
include 25 curves corresponding to 5 choices of the neutron $^1$S$_0$
and of the proton $^1$S$_0$ gaps covering the range of predictions
about the sizes of these gaps.  
\label{Fig:plot_cool_lum_data} }
\end{figure}

The observational data against which we will compare our results are
displayed in Figure~\ref{Fig:plot_data} and described in Appendix A.
An extensive set of full neutron star cooling histories employing the
minimal cooling scenario are displayed in
Figure~\ref{Fig:plot_cool_lum_data} from which the effects of the
vector channel suppression of Cooper-pair emission can be discerned.
A total of 300 cooling histories are shown and superposed with the
data from Figure~\ref{Fig:plot_data}.  Comparing the left and right
column panels of Figure~\ref{Fig:plot_cool_lum_data}, it is clear that
for times later than the crust thermal relaxation time ($t\simgreater
100$ years) the effects of suppressing Cooper pair emission from the
vector channel are very small.  Therefore, the major conclusions that
were reached in Paper I concerning minimal cooling remain valid.

In order to highlight the scope and limitations of the minimal cooling
paradigm, and to refine its implications, we recall its main
conclusions as stated in Paper I.  Given the uncertainties about the
major physical ingredients, namely, the sizes of the various pairing
gaps and the chemical composition of the envelope, the cooling curves
of the minimal model appear to be consistent with data of nearly all
neutron stars with observed thermal emission, as long as the coldest
possible trajectories under the minimal cooling paradigm occur in some
cases.  The compatibility of nearly all neutron stars with minimal
cooling requires strong conditions on the size of the neutron
$^3$P$_2$ gap and the chemical composition of the stellar envelopes,
as discussed below.

A few exceptions exist, notably the pulsar in CTA~1, together with the
unobserved emissions from several young supernova remnants that might
be reasonably expected to contain neutron stars, as they evidently lie
below any of the minimal cooling histories displayed.  These few objects, if
confirmed, provide us with the most compelling evidence in favor of
the occurrence of enhanced neutrino emission beyond the minimal
cooling paradigm.

In the case that the $T_c$ curve crosses the compatibility band as in
our models ``a'' and ``T'', agreement with data can be achieved if the
warmest stars, such as the pulsar in Puppis A and PSR 1E1207-52, have
light element envelopes, while the coldest ones, like PSR J1119-6127,
the pulsars in 3C58 and Vela, and PSR J1124-5916, have heavy element
envelopes. This agreement is lost, however, if the value of \tcmin~is
too large, i.e., greater than about $0.2 \times 10^{9}$ K (see
equation~\ref{Eq:Tcmin}), as in our models ``b'' and ``c''.  In the
extreme case that the neutron $^3$P$_2$ gap is vanishingly small and
also that all observed young cooling neutron stars have light element
envelopes, then nearly all of them, with the possible exception of PSR
B0538+2817, are observed to be too cold to be compatible with minimal
cooling predictions.  In the less extreme possibility of a
heterogeneity in chemical composition and a vanishingly small neutron
$^3$P$_2$ gap, we still find that more than half (seven out of twelve)
of the observed young cooling neutron stars are too cold to be
compatible with minimal cooling.  (Notice that among the remaining
five, out of twelve stars, the compact objects in Cas A and the Crab
still have only upper limits.)  If these conditions on the $T_c$ curve
are not satisfied for a particular model of superfluidity in dense
matter, then that model also requires enhanced cooling beyond the
minimal cooling paradigm.  These results highlight the importance of
the n $^3$P$_2$ gap in more precise terms than discussed in Paper I.

Our conclusion regarding the need for heterogenity in the
chemical composition of the atmosphere is consistent with the results
of \citet{Kam06}, who had to employ both light and heavy element
atmospheres in their cooling models to match the data of most stars.

That it is apparently possible to explain the majority of
thermally-emitting neutron stars with the minimal cooling paradigm led
\cite{Klahn06} to propose the so-called direct Urca constraint which
would effectively place limits on the density dependence of the
nuclear symmetry energy.  This follows from the fact that higher mass
stars have larger central densities, so that the few stars that appear
to be too cold could have larger than average neutron star masses. The
nearly two dozen well-measured neutron star masses are mostly
concentrated in the range 1.25--1.5 M$_\odot$~\citep{LP05}, although
there are indications that some neutron stars might extend beyond the
upper end of this range. Mass-dependent cooling in the context of
direct Urca processes was suggested by \cite{LPPH91}, and has been
exploited by \cite{PA92} and \cite{YP04} to account for the gross
features of cooling data. If rapid cooling is, in fact, due to the
onset of the direct Urca process involving nucleons, as opposed to
hyperons, kaon condensates, or deconfined quarks, a constraint on the
density dependence of the nuclear symmetry energy could therefore
result. The nuclear symmetry energy would have to grow relatively
slowly with density so that the threshold density for the direct Urca
process to occur exceeds the central density of most
neutron stars, few of which could be very massive.
The inferred constraints on the symmetry energy may also depend
significantly on the presence of quartic terms in the symmetry 
energy~\citep{Ste06}.

However, if any of the assumptions in the above logical sequence are
violated, the direct Urca constraint becomes invalid.  Envelope
compositions and the neutron $^3$P$_2$ gap might be outside the
required ranges.  Perhaps even in massive neutron stars cooling due to
the direct Urca process is abated due to superfluid-induced
suppression.  Furthermore, in the exceptional cases in which rapid
cooling is suspected, it will be difficult to prove that the nucleon
direct Urca is responsible as all proposed rapid cooling emission
rates are orders of magnitude greater than modified Urca or
Cooper-pair emission rates.  It should also be noted that there are at
present no reliable mass measurements of any thermally-detected
cooling neutron stars, so that mass information deduced from binary
pulsars might not be relevant.

\section{SUMMARY AND CONCLUSIONS}

Neutrinos emitted in the continual Cooper-pair breaking and formation
(PBF) processes that occur near the critical temperatures of
superfluid neutrons and superconducting protons in neutron star
interiors are an integral part of the minimal cooling paradigm (Paper
I). These processes were not generally included in the traditional
standard cooling scenario, but are nevertheless expected to occur
whether or not rapid cooling occurs.  If all neutron star cooling data
are found to be consistent with the predictions of the minimal cooling
model, (in which rapid cooling by direct Urca processes occurring
through reactions involving nucleons, hyperons, quarks, or Bose
Condensates were eliminated by design) then there is no reason to
invoke rapid cooling.  Either rapid cooling does not occur or it is
suppressed by superfluidity.  In either case, constraints about the
properties of dense matter ensue. 

However, beginning with \cite{LP06a}, several authors
\citep{SMS07,EV08,L08,SR08} have recently established that the
PBF neutrino emissivity from the vector channel is
suppressed by a factor of $\sim 10^{-3}$ from the original estimates
of \cite{FRS76}, who overlooked the conservation of vector current in
weak interactions.  In view of this development, we have performed
extensive calculations to assess the effects of the severely reduced
rates in the vector channel on the long-term cooling of neutron stars
incorporating the revised PBF rates.

Our analysis leads us to conclude that the long-term cooling of
neutron stars as envisaged in the minimal cooling paradigm is not
significantly affected by the proposed large reductions in the
vector channel of the PBF neutrino emissivities. The reason is that
Cooper-pair emission occurs through both vector and axial
channels, and the axial part of the Cooper-pair emission is barely
affected.  The axial channel of the PBF emissivites controls the
long-term cooling, whereas the vector channel emissivities are
important for the cooling behavior of neutron stars with ages less
than a few hundreds of years for which observational data are not yet
available.  With the exception of few candidates (e.g., CTA 1 and
perhaps unobserved stars in other remnants), the minimal cooling
paradigm is consistent with observations of neutron star temperatures
and ages, but, as noted in Paper I, only if certain combinations
of light- or heavy-element envelopes and sizes of neutron triplet
pairing gaps in the star's core exist.

For the minimal cooling model to be consistent with data, it is
necessary for the most efficient cooling possible to occur.  Therefore,
our analysis places a stringent requirement on the critial temperature
for neutron superfluid pairing in the triplet channel in the core of
the star.  It is required that the $T_c$ curve crosses the
compatibility band described in Figure~\ref{Fig:Tc}, and therefore the
triplet gap must both become larger than $T_c=0.5\times10^9$ K at some
moderately large density (but less than the central density of the
neutron star), and also become smaller than $T_c=0.2\times10^9$ K
(either at high density, i.e., in the center of the star, or 
at low density, i.e. in the outermost part of the core).
If the gap does not fulfill these conditions,
then observations point to the occurence of enhanced cooling in at
least half of the young isolated cooling neutron stars observed so
far.

\acknowledgements 

The authors acknowledge discussions with Edward Brown and Sanjay
Reddy.  DP acknowledges support from UNAM-DGAPA grants of the PAPIIT
program (\#IN119306 and \#IN122609). JML and MP acknowledge research support from the
U.S. DOE grants DE-AC02-87ER40317 and DE-FG02-93ER-40756,
respectively.  JML would also like to acknowledge support from a
Glidden Visiting Professorship Award at Ohio University.  AWS is
supported by the Joint Institute for Nuclear Astrophysics under
NSF-PFC grant PHYS 08-22648, by NASA ATFP grant NNX08AG76G, and by the
NSF under grant number PHY-0456903.  The hospitality of the National
Institute of Nuclear Theory, University of Washington, Seattle, USA,
where this work was begun is acknowledged.

\appendix

\section{Observational Data}
\label{App:data}

\bigskip

The observational data plotted in Figures~\ref{Fig:plot_data} and
\ref{Fig:plot_cool_lum_data} in the text deserve some comments. We
have considered the same objects and data as reported in Paper I, but
with modifications for the first five objects discussed below. We have
also included some additional objects (see the last four objects
below) for which data has become available since Paper I was
published. Salient aspects of the modifications and new data are
summarized here.

In many cases, the only way in which the ages of the objects can be
estimated is by their spin-down ages $2P/\dot P$. However, there are
many stars for which kinematic ages have also been obtained. Comparing
kinematic and spin-down ages for the same objects often reveals
discrepancies of factors of three or more in both directions.  For
example, SNR N157B has a determined age of less than 2000 years but a
pulsar spin-down time of 5000 years, and PSR B1757-24 has an estimated
age less than 39000 years and a spin-down age of 16000 years. For this
reason, we have attached nominal error bars of a factor of three to
those objects for which we only have spin-down information.  To ignore
this uncertainty could be misleading.

\subsection{ PSR RX J0205+6449 in the nebula 3C58} 

The supernova remnant 3C58 has been associated with the historical
supernova SN 1181 \citep{SG99} providing us with a pulsar age of 828
years.  However, this proposed association has been challenged in
recent years.  Measurements of the remnant's expansion velocity
compared to its size (at an assumed distance of 3.2 kpc) require an
older origin. Optical observations of filament expansion (angular
expansion) and filament line width (radial expansion) imply an age of
3000-4000 yrs and 2700-3500 yrs, respectively \citep{FRS08}, whereas radio
measurements of the filament expansion would imply an age of 4500-7000
yrs \citep{B06}.  Moreover, modeling of the energetics of the pulsar
wind nebula and its expansion need an age of 2000-3000 yrs
\citep{C04,C05} or even 2000-5000 yrs from the recent X-ray
observations of \cite{GHN07}.  These ages are more in accord with the
pulsar spin-down age of 5400 yrs.  We will, henceforth, adopt a
conservative age of 2000-7000 yrs, keeping in mind the possible
SN 1181 association and the corresponding age of 828 yrs.

\subsection{ PSR B0833-45 in the SNR Vela}

The kinematic age of the SNR we adopted in Paper I was taken from
\cite{AET95} who employed a distance of 500 pc (the PSR dispersion
measure distance). Correcting this distance to 300 pcs (the VLBI
parallax distance of \citealt{Detal2003}) gives us an age of
5400-16000 yrs as emphasized by \cite{Tetal09}.

\subsection{ RX J0007.0+7302 in the SNR CTA1}

We notice that the {\em Fermi} Gamma-Ray space telescope has detected
pulsations at 315 ms from this compact source \citep{Aetal08} and
measured its spin-down age as $1.4 \times 10^4$ yrs, in agreement with
the estimated age of the remnant, $1.4 - 2.2 \times 10^4$ yrs.

\subsection{PSR 1055-52}

This pulsar's spin-down age is $10^{5.73}$ yrs instead of $10^{5.43}$
yrs which was erroneously adopted in Paper I.

\subsection{PSR 1E1207-52}

A lower limit on the spin-down age of this pulsar has been found to be
27 Myrs \citep{GH07} while the associated supernova remnant kinematic
age is about 7000 yrs (within a factor 3). As in Paper I, we use the
kinematic age from the SNR.

\subsection{PSR B2334+61 in the SNR G114.3+0.3}

Surface thermal emission has been detected in an XMM-Newton
observation \citep{McG06}.  The distance estimate from the pulsar
dispersion measure is $3.1^{+0.2}_{-1.0}$ kpc, but HI absorption
\citep{YUUK04} indicates a much shorter distance of less than 1 kpc.
Fits of the spectrum with a magnetized hydrogen neutron star
atmosphere (NSA) gives $T^{NSA}_\infty = 0.65 \times 10^6$ K and
$R^{NSA}_\infty \sim 13$ km, whereas a blackbody (BB) fit gives
$T^{BB}_\infty = 1.6 \times 10^6$ K and $R^{BB}_\infty \sim 1.6$
km. In both cases, a distance of 3.1 kpc was assumed.  The hydrogen
atmosphere model seems to give better fits, but the BB fit would be
preferred in case of the smaller distance.  However, the uncertainty
on the luminosity is dominated by the uncertainty in the distance and
with the above values, we obtain $L_\infty = 0.5 \times 10^{32} - 5
\times 10^{32}$ erg s$^{-1}$ using the NSA values, whereas the BB
values give an $L_\infty$ range included within the NSA results.  The
age of the SNR, a few times $10^4$ yrs, is uncertain as is the
association with the PSR whose spin-down age is $4 \times 10^4$
yrs. We will adopt an age of $(\frac{1}{3} - 3) \times t_{sd}$, i.e.,
with the same uncertainty factor as we did in Paper I.

\subsection{PSR J1119-6127}

This high-field pulsar shows a thermal spectrum which when fitted by a
blackbody spectrum gives $L^{BB}_\infty = (1.35 - 2.25) \times
10^{33}$ erg s$^{-1}$.  NSA fits yield $L^{NSA}_\infty = (1.05 - 2.05)
\times 10^{33}$ erg s$^{-1}$ \citep{SHK08}.  The NSA fit is compatible
with emission from the entire neutron star surface whereas the BB fit
implies a much smaller radius ($\sim 3.5$ km) at an estimated distance
of 8.5 kpc.  However, the distance is uncertain and the exceptionally
high pulsed fraction ($\sim$ 75 \%) \citep{Getal05} may favor a small
thermally emitting region.  Given this and because $L^{BB}_\infty$ and
$L^{NSA}_\infty$ are similar, we adopt $L^{BB}_\infty = (1.05 - 2.25)
\times 10^{33}$ erg s$^{-1}$ for the thermal luminosity.  The
spin-down age is $t_{sd} = 1600$ yrs.  As the measured braking index
$n=2.9$ \citep{Cetal00} is compatible with magneto-dipolar breaking, and
the characteristics of the associated SNR and the Pulsar Wind Nebula
(PWN) are compatible with an age close to $t_{sd}$ \citep{SHK08}, we
adopt an age estimate of $(0.5 - 2) \times t_{sd}$.

\subsection{ PSR B0531+21 in the Crab nebula}

\noindent Phase resolved X-ray spectroscopy \citep{Wetal04} gives the most
stringent upper limit on the surface thermal luminosity, $L_\infty <
2.8 \times 10^{34}$ erg s$^{-1}$.  However, there is no evidence of a
thermal component in the spectrum which is completely dominated by
magnetospheric emission.

\subsection{ SNRs G065.3+5.7 and G074.0-8.5}
\label{Sec:A8}

\noindent In their extensive search for compact objects in nearby
SNRs, \cite{Ketal06} provide six new upper
limits on the thermal luminosity of possibly existing (but undetected)
compact stars. Four of these limits are quite high so we will retain
only the two lowest ones: $L_\infty < 10^{32}$ erg s$^{-1}$ for SNR
G065.3+5.7 (Cygnus Loop) and $L_\infty < 0.7 \times 10^{32}$ erg
s$^{-1}$ for G074.0-8.5.  The estimated ages of the remnants are $(2 -
2.5) \times 10^4$ yrs and $(0.6 - 1.2) \times 10^4$ yrs, respectively.



\end{document}